\documentclass{moriond}

\usepackage{graphicx}

\graphicspath{{./fig/}}

\usepackage{amsmath}

\def\half{{\textstyle{1\over2}}}

\def\be{\begin{equation}}
\def\ee{\end{equation}}
\def\bea{\begin{eqnarray}}
\def\eea{\end{eqnarray}}
\usepackage{amsmath}

\newcommand{\Photo}{\includegraphics[height=25mm]{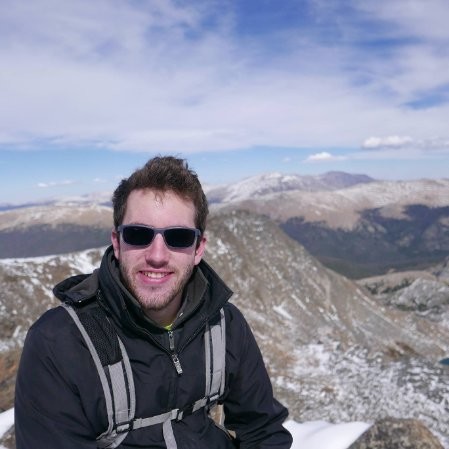}}

\begin{document}
\vspace*{4cm}
\title{Gravitational waves from MHD turbulence at the QCD phase transition
as a source for Pulsar Timing Arrays}

\author{ A. Roper Pol}

\address{Université Paris Cité, CNRS, Astroparticule et Cosmologie, F-75013 Paris, France\\
School of Natural Sciences and Medicine, Ilia State University, 
GE-0194 Tbilisi, Georgia}

\maketitle
\abstracts{
We propose that the recent observations reported by the different Pulsar Timing Array (PTA)
collaborations (i.e.~IPTA, EPTA, PPTA, and NANOGrav) of a common process over several pulsars
 could correspond to a
stochastic gravitational wave background (SGWB) produced by turbulent sources in the early
universe, in particular due to the magnetohydrodynamic (MHD) turbulence induced by primordial
magnetic fields.
I discuss recent results of numerical simulations of MHD turbulence
and present an analytical template of the SGWB
validated by the simulations.
We use this template to constrain the magnetic field 
parameters using the results reported by the PTA collaborations.
Finally, we compare the constraints on the primordial magnetic fields
obtained from PTA with those from blazar signals observed by Fermi Large Area Telescope (LAT),
from ultra high-energy cosmic rays, and from the cosmic microwave background.
We show that a non-helical primordial magnetic field produced at the scale
of the quantum chromodynamics phase transition is compatible with such constraints and it could additionally provide with a magnetic field at recombination that
would help to alleviate the Hubble tension.\\
{\bf \em Contribution to the 2022 Gravitation session of the 56th Rencontres de Moriond.}}

\section{Introduction}

The field of gravitational astronomy is near its birth and has already led to
astonishing discoveries.
Starting in 2015 with the first detected gravitational wave event GW150914,\,\cite{LIGOScientific:2016aoc} and the first multi-messenger detection
GW170817 in 2017,\,\cite{LIGOScientific:2017vwq} the LIGO-Virgo collaboration has recently released the second half of the third observing run O3b, with a total of 90 observed events.\,\cite{LIGOScientific:2021djp}
Many more are expected to come in the next few years
and decades.
In particular, magnetic fields and bulk fluid motions in the early universe can
source a stochastic gravitational wave background (SGWB) of cosmological origin that can be generated, for example, during cosmological phase transitions.\,\cite{Caprini:2018mtu}
Depending on the energy scale at which the SGWB is produced, it will be present
in a different range of frequencies within the gravitational spectrum.
For example, a signal produced at the QCD phase transition with an
energy scale $T_* \sim 100$\,MeV could produce a SGWB at
frequencies of a few nanohertz.\,\cite{Deryagin:1986qq}
It is precisely in the 1--100 nHz regime where pulsar timing arrays (PTA) seek to detect GWs by monitoring the
time-delay signals of an array of millisecond pulsars.
Recently, the North American Nanohertz Observatory for
Gravitational Waves (NANOGrav), after 12.5 years of observations,
has reported a common-spectrum process, 
although the statistical significance of a quadrupolar correlation (i.e.~following the Hellings-Downs curve as expected for GW signals\,\,\cite{Hellings:1983fr})
is still not conclusive.\,\cite{Arzoumanian:2020vkk}
Similar results from the Parkes Pulsar Timing Array (PPTA),\,\cite{Goncharov:2021oub} the
European Pulsar Timing Array (EPTA),\,\cite{Chen:2021rqp} and the International
Pulsar Timing Array (IPTA) collaborations\,\,\cite{Antoniadis:2022pcn} followed, using data from pulsars
that span a total duration of 15, 24, and 31 years, respectively.
Different sources of GWs that would yield a SGWB compatible with the common-process
observed by the PTA collaborations have been proposed.
The most studied source is astrophysical and corresponds to the SGWB produced by
a population of merging supermassive black hole binaries.\,\cite{Haehnelt:1994wt}
Alternatively,
cosmological sources have been proposed: inflation,\,\cite{Vagnozzi:2020gtf} cosmic strings
and domain walls,\,\cite{Ellis:2020ena} primordial black holes,\,\cite{Vaskonen:2020lbd} supercooled and dark phase transitions,\,\cite{Nakai:2020oit} the QCD phase
transition, and primordial magnetic fields.\,\cite{Neronov:2020qrl}
I present hereby the latter scenario,
where a primordial magnetic field produced or present during the
QCD phase transition yields a SGWB that is compatible with the observations reported
by the PTA collaborations.\,\cite{Neronov:2020qrl,Brandenburg:2021tmp,RoperPol:2022iel}
In particular, I present the results of Roper Pol et al.~2022.\,\cite{RoperPol:2022iel}

\section{MHD turbulence and gravitational wave production}

In the presence of magnetic fields, due to the high-conductivity
in the early universe, the plasma velocity field is highly coupled to the
primordial magnetic field, leading inevitably to the development of magnetohydrodynamic (MHD)
turbulence.\,\cite{Ahonen:1996nq}
On the other hand, the anisotropic stresses due to the velocity and
magnetic fields lead to the production of GWs.
To compute the resulting SGWB, one needs to solve the full system
of MHD equations.
We have performed such numerical simulations in a recent publication\,\,\cite{RoperPol:2022iel} using the
{\sc Pencil Code}\,\,\cite{PencilCode:2020eyn} and
following a similar setup than previous
simulations of MHD turbulence production of GWs from cosmological phase transitions.\,\cite{Brandenburg:2021tmp,Pol:2018pao,Pol:2019yex,Kahniashvili:2020jgm,Brandenburg:2021bvg,RoperPol:2021xnd}
The numerical setup and methodology,
and some of these results, in particular
focusing on the electroweak phase transition
and the potential detectability of polarization in the SGWB,
were presented during the Gravitation session of the
55th Rencontres de Moriond.\,\cite{RoperPol:2021gjc}

The tensor-mode perturbations $h_{ij}^{\rm phys}$ above the Friedmann-Lema\^itre-Robertson-Walker background metric tensor
are described by the GW equation.
During the radiation-dominated era, it reads
\begin{gather}
\left(\partial_t^2 - \nabla^2\right) h_{ij} = 6\,\Pi_{ij}/t,
\label{GW_eq_norm}
\end{gather}
for scaled strains $h_{ij}=ah_{ij}^{\rm phys}$, comoving coordinates and stress tensor components,
and conformal time.
We have used a normalization appropriate for numerical simulations,\,\cite{Pol:2018pao,Pol:2019yex}
and $\tilde \Pi_{ij} = \Lambda_{ijlm} \tilde T_{lm}$ (where a tilde indicates
Fourier space) corresponds to the traceless-transverse projection of
the stress-energy tensor produced by the MHD turbulence,
\begin{equation}
T_{ij}=(p/c^2 + \rho) \gamma^2 u_i u_j - B_i B_j + \delta_{ij} B^2/2,
\label{eq:Tmunu}
\end{equation}
where $\Lambda_{ijlm} =P_{il} P_{jm} - \half P_{ij} P_{lm}$
and $P_{ij} = \delta_{ij} - 
\hat k_i \hat k_j$ are projection operators.

\subsection{Numerical results}

We solve the MHD equations
numerically and construct the
stress tensor $\Pi_{ij}$; see eq.~\eqref{eq:Tmunu}.
Then, we compute the 
resulting GW radiation solving eq.~\eqref{GW_eq_norm}.
For this purpose, we use the open-source {\sc Pencil Code}.

\begin{figure}
\begin{minipage}{0.5\linewidth}
\centerline{\includegraphics[width=\linewidth]{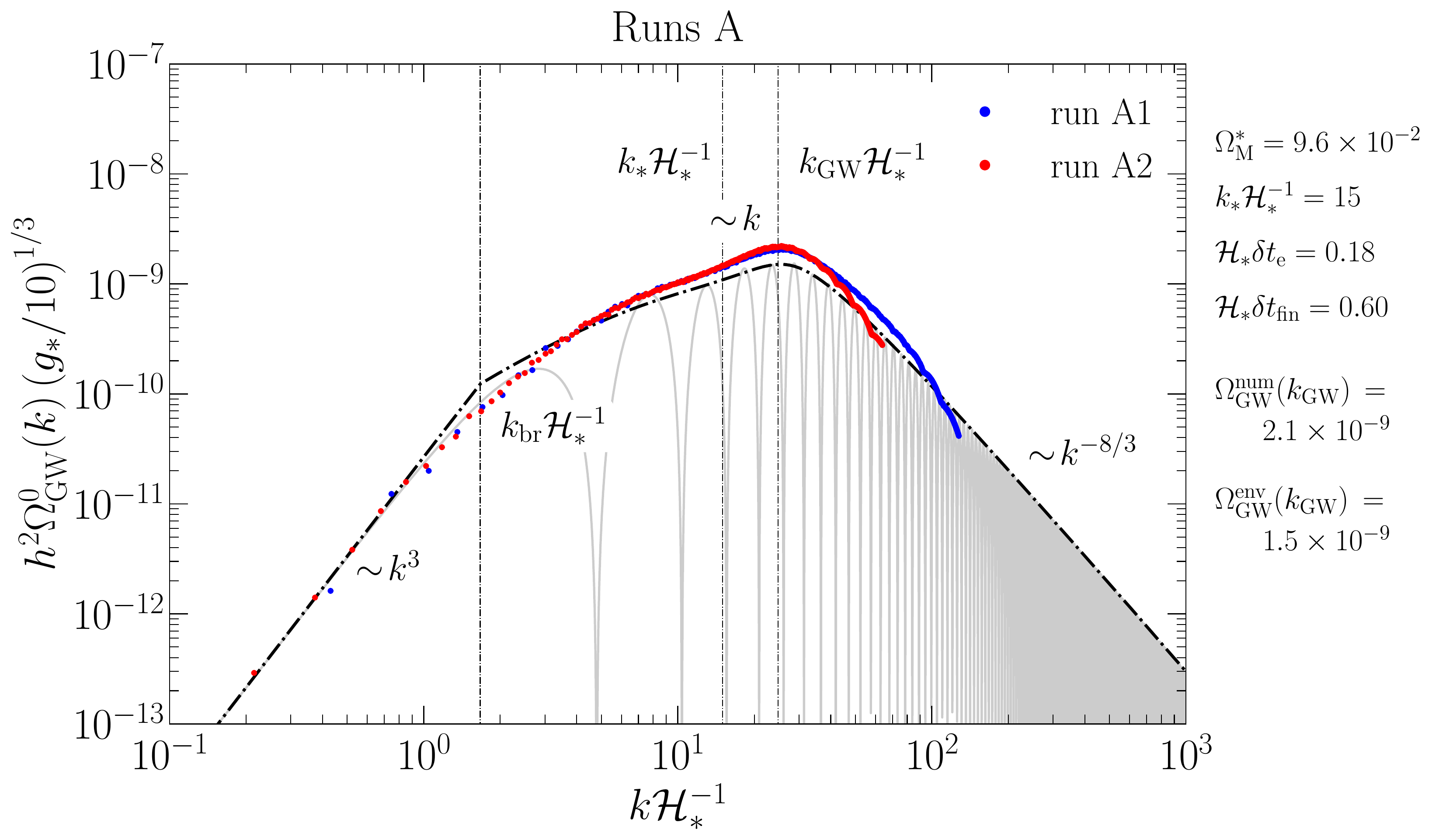}}
\end{minipage}
\hfill
\begin{minipage}{0.5\linewidth}
\centerline{\includegraphics[width=\linewidth]{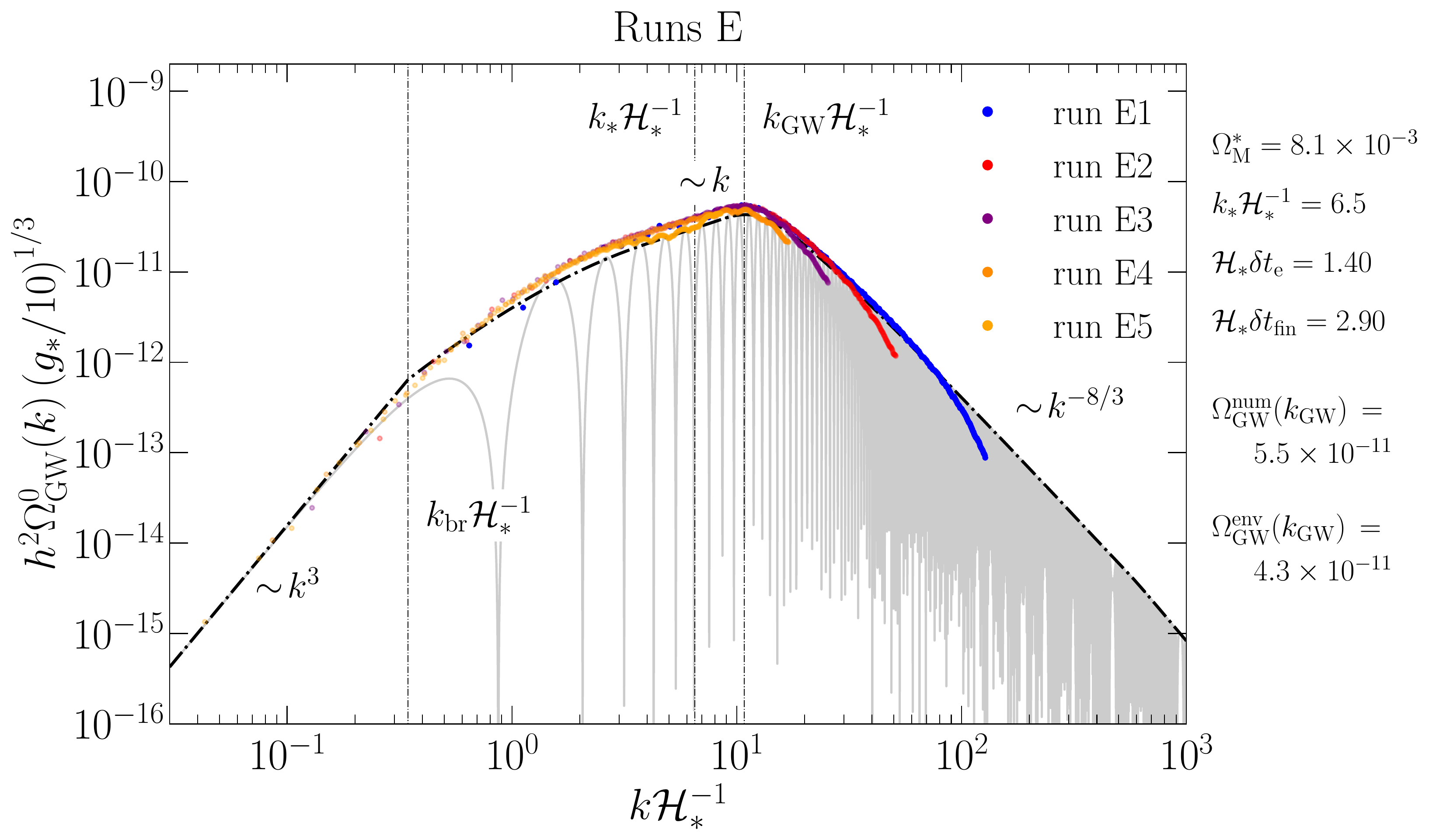}}
\end{minipage}
\hfill
\caption[]{SGWB spectrum produced by MHD turbulence driven by non-helical 
magnetic fields at the QCD phase transition, computed
numerically using the {\sc Pencil Code} (colored dots), and obtained
from an analytical template (solid lines) and its envelope (dash-dotted lines).
The left panel shows a series of runs with a short eddy turnover time
(18\% of one Hubble time) and the right panel shows a series
of runs
with a long eddy turnover time (1.4 Hubble times).
The specific parameters of each run and the results can be
found in Roper Pol et al. 2022.\,\cite{RoperPol:2022iel}}
\label{fig:OmegaGW_num}
\end{figure}

We set as initial condition of the simulations a 
fully-developed non-helical MHD spectrum for the magnetic field, and zero 
initial bulk velocity, present at the QCD phase transition (in general, at
an energy scale $T_*$) with a characteristic scale $k_*$, given as a fraction of the
Hubble scale, and a characteristic strength $\Omega_{\rm M}^*$,
given as a fraction of the critical energy density.
We chose this initial condition for two main reasons.
In first place, it is conservative: whatever the initial generation mechanism, the magnetic field is expected to enter a phase of fully developed and freely decaying turbulence.\,\cite{Ahonen:1996nq}
Any initial phase of magnetic field growth would increase the GW production.\,\cite{Pol:2019yex,Kahniashvili:2020jgm,RoperPol:2021xnd}
Secondly, simple initial conditions make it easier to build an analytical description of the simulation outcome.  
Since we want to model magnetically driven turbulence, we also neglect the presence of initial bulk velocity for simplicity.
The resulting SGWBs computed from the numerical simulations
are shown in
figure~\ref{fig:OmegaGW_num}.


\subsection{Analytical SGWB template}

In general, the simulations show that the SGWB evolves with
a characteristic time scale $\delta t_{\rm GW} \sim 1/k_*$,
while the magnetic field evolution is determined by the eddy
turnover time, $\delta t_{\rm e} = (v_{\rm A} k_*)^{-1}$,
where $v_{\rm A}$ is the Alfv\'en speed.
Hence, due to causality, $\delta t_{\rm GW}/\delta t_{\rm e}
\sim v_{\rm A} < 1$.
This reasoning motivates the assumption that the anisotropic 
stresses are constant in time, which simplifies the solution
to the GW equation; see eq.~\eqref{fig:OmegaGW_num}.
In particular, the envelope of the resulting SGWB spectrum,
derived in Roper Pol et al. 2022,\,\cite{RoperPol:2022iel}
is
\begin{align}
    \Omega_{\rm GW} (k,  t_*) \approx   3 \, \biggl(\frac{k}{k_*}\biggr)^3 \, {\Omega_{\rm M}^*}^{2} \frac{{\cal C} (\alpha)}{{\cal A}^{2}(\alpha)} 
     \,\,p_\Pi \biggl(\frac{k}{k_*}\biggr)
   \left\{
    \begin{array}{lr}
       \ln^2 [1 + {\cal H}_* \delta t_{\rm fin}]  &
       \text{ if } k\, \delta t_{\rm fin} < 1, \\
        \ln^2 [1 + (k/{\cal H}_*)^{-1}]
       & \text{ if }  k\, \delta t_{\rm fin} \geq 1.
    \end{array}
    \right.
    \label{OmGW_envelope}
\end{align}
The SGWB shows a dependence with the square of the magnetic
field strength $\Omega_{\rm M}^*$ and the magnetic spectral
shape via $p_\Pi (k/k_*)$.
In this work, we consider an initial magnetic spectrum given as a
smoothed broken power law, following a Batchelor spectrum $\sim\!k^4$ at large scales, and a Kolmogorov spectrum $\sim\!k^{-5/3}$ at small scales,
with $\alpha = 2$ determining the smoothness of the transition; see eq.~(6) of Roper Pol et al. 2022.\,\cite{RoperPol:2022iel}
The parameters ${\cal C}$ and ${\cal A}$ are numerical values that
depend on $\alpha$, and $p_\Pi$ is given by the projection of the 
convolution of the magnetic spectrum for Gaussian magnetic fields; see eq.~(11) of Roper Pol et al. 2022.\,\cite{RoperPol:2022iel}
In this expression, we have considered an effective
duration of the turbulence sourcing $\delta t_{\rm fin}$ that can
not be predicted by the analytical model.
We expect that this duration is proportional to the time scale of the
magnetic evolution, i.e.~the eddy turnover time.
Hence, we use the results from the numerical simulations to provide the
empirical fit $\delta t_{\rm fin} = 0.184 {\cal H}_*^{-1} + 1.937 \,\delta t_{\rm e}.$
Figure~\ref{fig:OmegaGW_num} compares the analytical model with the
results from the numerical simulations, and shows that the envelope of
the analytical template is an accurate representation of the numerical
results.\,\cite{RoperPol:2022iel}

\section{Constraints on the magnetic field}

\begin{figure}
\centerline{\includegraphics[width=.7\linewidth]{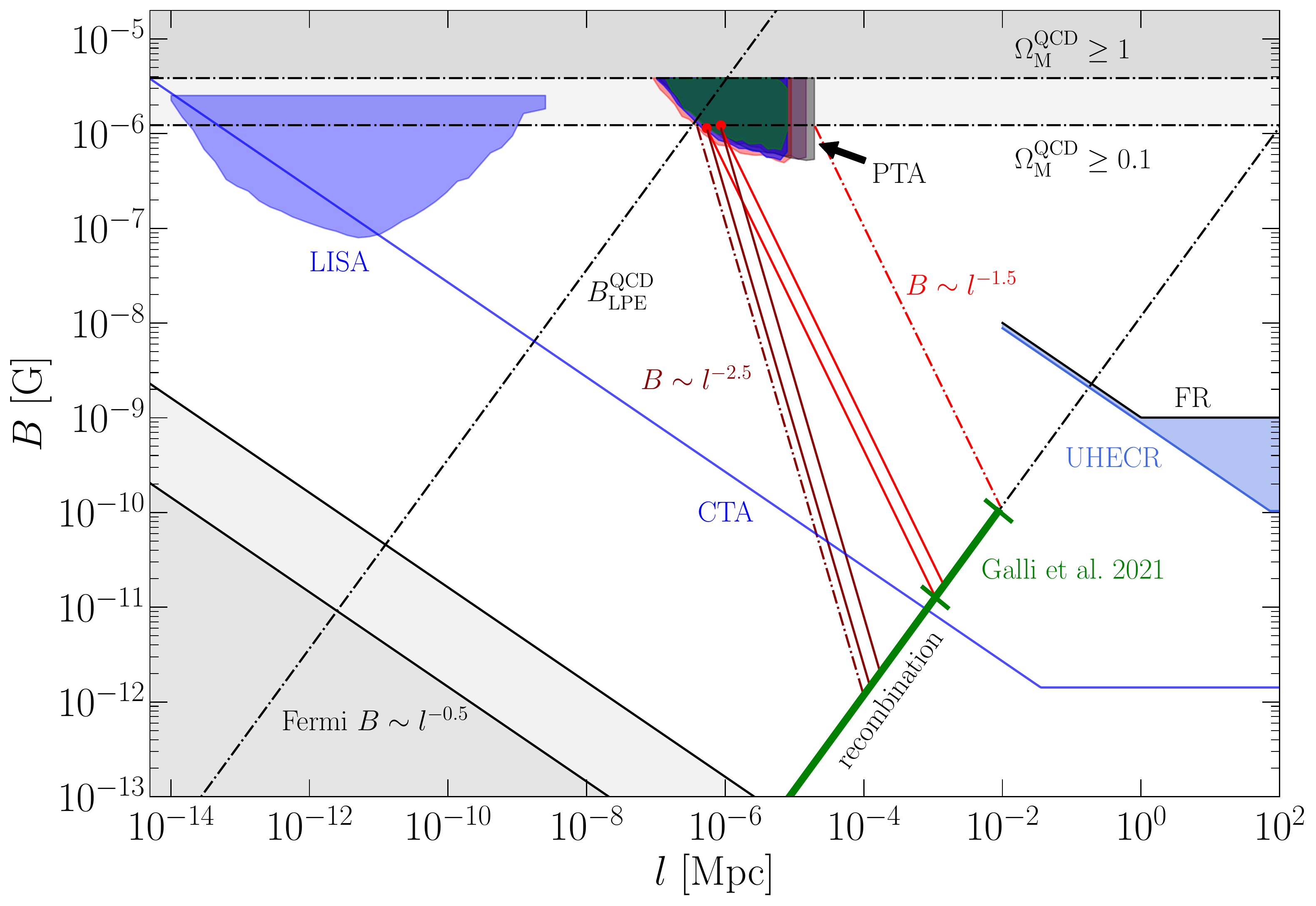}}
\caption[]{Magnetic field amplitude $B$ and length at the spectral peak
$l$ given as present time observables (i.e.~in comoving units)
compatible with
the observations from PTA compared to the different constraints
on primordial magnetic fields.
The red and brown lines indicate possible
evolutionary paths of the
primordial magnetic field from its generation
down to recombination. For details, see
Fig.~7's caption in Roper Pol et al. 2022.\,\cite{RoperPol:2022iel}}
\label{fig:OmGW_PTA}
\end{figure}

Once that we have derived an analytical template of the SGWB produced
by MHD turbulence due to the presence of a non-helical magnetic field in
the early universe, which has been validated by the simulations, we can
compare the resulting SGWB to the observations reported by the different
PTA collaborations,\,\cite{Arzoumanian:2020vkk,Goncharov:2021oub,Chen:2021rqp,Antoniadis:2022pcn}
thereby inferring the range of parameters $T_*$, $k_*$, and $\Omega_{\rm M}^*$ that could account for the PTA results.
The details and results of this analysis are described in Roper Pol et al. 2022.\,\cite{RoperPol:2022iel}
Figure~\ref{fig:OmGW_PTA} shows the range of parameters
$k_*$ and $\Omega_{\rm M}^*$, for the compatible range of energy scales $T_*$,
that are compatible with the results of the different PTA collaborations.
They are shown in terms of the magnetic
field amplitude $B$ (in Gauss) and the length scale of the spectral
peak $l$ (in Mpc), as observables at present time.\,\cite{RoperPol:2022iel}
Such magnetic fields will continue to evolve following MHD turbulent free
decay down to recombination, when the magnetic fields become frozen-in
and only evolve following the expansion of
the universe afterwards.\,\cite{Durrer:2013pga}
Figure~\ref{fig:OmGW_PTA} shows that the magnetic fields compatible with the
observations from the PTA collaborations are also compatible with
constraints on large-scale magnetic fields from the Fermi
collaboration,\,\cite{Neronov:1900zz}
constraints from ultra-high-energy cosmic rays (UHECR),\,\cite{TelescopeArray:2021dfb} constraints
from Faraday Rotation (FR),\,\cite{Pshirkov:2015tua}
and constraints from the cosmic microwave background (CMB).\,\cite{Galli:2021mxk}
In fact, the resulting primordial magnetic field from the PTA observations
is also compatible with the constraints proposed by Jedamzik \& Pogosian 2020,\,\cite{Jedamzik:2020krr} updated recently by Galli et al.
2022,\,\cite{Galli:2021mxk} which allow a CMB value of the
current Hubble rate $H_0 \approx 70$\,km\,s$^{-1}$\,Mpc$^{-1}$, hence
relieving the Hubble tension.

\section{Conclusions}

I have presented recent results on the SGWB produced by MHD turbulence
driven by primordial magnetic fields from the radiation-dominated
early universe.
New MHD numerical simulations allowed us to study the relevant dynamical range of the GW spectrum
and to validate an analytical template of the SGWB that holds under
the assumption of constant-in-time anisotropic stresses.
Whether the PTA observations are confirmed to correspond to a SGWB (i.e.~if the correlation between pulsars is confirmed to follow the Hellings-Downs 
curve) in the future, they could potentially correspond to a SGWB produced
by primordial magnetic fields with a magnetic energy density
of at least $1\%$ the radiation energy density and a characteristic length scale within 10\% of the horizon scale.
The energy scale is constrained to be between 1 and 200 MeV and hence,
the signal could correspond to a primordial magnetic field produced during the
QCD phase transition.
The analytical model developed, validated by the simulations, allowed us to
predict where the position of the break from $f^3$ to $f$ appears within
the SGWB.
This break, if observed in future PTA data, could help to elucidate
among an astrophysical or cosmological origin of the SGWB signal.\,\cite{RoperPol:2022iel}

\section*{Acknowledgments}

I would like to acknowledge my collaborators C. Caprini,
A. Neronov and D. Semikoz for their fruitful contributions to this work.
Support through the French National Research Agency (ANR) project MMUniverse (ANR-19-CE31-0020) and the Shota Rustaveli National Science Fundation of Georgia (grant FR/18-1462)
are gratefully acknowledged.
I thank the {\em Rencontres de Moriond} organizers to give me the chance
to present this work and to the rest of participants 
for providing a great environment and interesting
discussions during the meeting.

\end{document}